
\magnification=1200
\vsize=7.5in
\hsize=5in
\tolerance 10000

\pageno=1
\def\ep{\delta}
\def\de{\epsilon}

\def\L{p_0}
\vskip .1in
\baselineskip 12pt plus 1pt minus 1pt
\smallskip
\centerline{NEGATIVE KINETIC ENERGY BETWEEN PAST AND FUTURE STATE
VECTORS{\footnote{*}{Talk presented at the Conference on Fundamental
Problems in Quantum Theory, Baltimore, MD, June 1994}}}
\vskip 24pt
\centerline{Daniel Rohrlich,
Yakir Aharonov,{\footnote{$\dagger$}{also \it
Department of Physics, University of South Carolina,
Columbia, SC  29208}} Sandu Popescu, Lev Vaidman}
\vskip 12pt
\centerline{\it Raymond and Beverly Sackler Faculty of Exact Sciences}
\centerline{\it School of Physics and Astronomy, Tel-Aviv University}
\centerline{\it Ramat-Aviv, Tel-Aviv 69978 Israel}
\vskip 1cm
{\bf Abstract.}
An analysis of errors in measurement yields new insight into
classically forbidden quantum processes. In addition to
``physical" values, a realistic measurement can yield ``unphysical"
values; we show that in {\it sequences} of measurements, the
``unphysical" values can
form a consistent pattern.  An experiment to isolate a particle in a
classically forbidden region obtains
a {negative} value for its kinetic energy.
It is the {\it weak value}
of kinetic energy between past and future state vectors.
\baselineskip 24pt plus 2pt minus 2pt
\goodbreak
\bigskip
{\bf 1. \quad Introduction.}
When the word ``quantum" first entered the language of physics, it meant a
restriction on possible values of energy; and
it is still axiomatic that the only
observable values of a physical quantity are the eigenvalues of a corresponding
quantized operator.
When we obtain values that are not eigenvalues,
we interpret them as errors. Still, measurements are uncertain in practice, and
can even yield classically forbidden, ``unphysical" values. We have uncovered
remarkable regularities in the way that ``unphysical" values can appear in
sequences of measurements, suggesting that these values may not be unphysical
at all.  In quantum theory, it seems, not only are physical quantities not
restricted: they can take values outside the classically allowed range.  Here
we discuss this new effect in the context of barrier
penetration by quantum particles.

     Barrier penetration, such as tunnelling into a potential wall, is
a classically forbidden quantum process. Quantum
particles can be found in regions where a classical particle could never go:
it would have negative kinetic energy.  But the
eigenvalues of kinetic energy cannot be negative.  How, then, can a quantum
particle ``tunnel"? The apparent paradox is resolved by noting that the wave
function of a tunnelling particle only partly overlaps the forbidden region,
while a particle found within the forbidden region may have taken enough energy
from the measuring probe to offset any energy deficit.
Nevertheless, actual measurements of kinetic energy can
yield negative values. Here, we present a
model experiment in which we measure the kinetic energy of a
bound particle to any desired precision. We then
attempt to localize the particle within the classically forbidden region. The
attempt rarely succeeds, but whenever it does, we find that the kinetic energy
measurements gave an ``unphysical" negative result; moreover, these results
cluster around the appropriate value, the difference between the total and the
potential energy. This consistency, which seems to come from nowhere -- a
background of errors -- suggests strongly that the notion of a quantum
observable is richer than generally realized.  Previous papers
making this suggestion analyze a measurement of spin$^1$
and a quantum time machine$^2$ as well as negative kinetic energy.$^{3-4}$
\goodbreak
{\bf 2. \quad Negative kinetic energy.}
\nobreak
Our example may be summarized as follows: we prepare a large
ensemble of particles bound in a potential well, in an eigenstate of energy,
and measure the kinetic energy of each particle to a given precision.
Then we
measure the position of each particle and select only those cases where the
particle is found within some region ``far enough" from the well -- with ``far
enough" depending on how precisely the kinetic energy was measured.  In almost
all such cases, we find that the measured kinetic energy values are
{\it negative} and  cluster around the
particular negative value appropriate to particles in the classically forbidden
region. Also, the spread of the clustering is the characteristic spread for
kinetic energy measurements with this device.

     We begin with a particle trapped in a potential well.  The Hamiltonian
is
$H={{p^2}/{2m}}+V(x)$,
with $V(x)=-V_0$ for  $\vert x\vert <a$ and $V(x)=0$ for $\vert x\vert >a$.
We prepare an ensemble of particles in the ground state,
with energy $E_0 <0$: $\vert \Psi_{in} \rangle =\vert E_{0}\rangle.$
Following von Neumann,$^5$ we model a measurement of kinetic energy with an
interaction Hamiltonian $H_{int}=g(t) P~{p^{2}\over 2m}$,
where $P$ is a canonical momentum conjugate to the position $Q$ of a pointer on
the measuring device. The
time-dependent coupling constant $g(t)$ is nonzero only for a short time
interval, and normalized so that
$\int g(t)dt=1$.
When the time interval is very short, we call the measurement impulsive. For an
impulsive measurement, $H_{int}$ dominates the Hamiltonians of the measured
system and the measuring device.  Then, since $\dot Q = {i \over \hbar}
[H_{int} ,Q]$, we obtain
for the operator $Q$
$$Q _{fin}-Q _{in}={p^{2}\over 2m}~~~~.
\eqno(1)$$
In an ideal measurement the position of the pointer is precisely
defined, and so we read a precise
value of kinetic energy.
But in practice, measurements involve uncertainty.  To model
a source of uncertainty, we take the initial state of the pointer to be
$$\Phi _{in} (Q) = (\de^2 \pi )^{-1/4} e^{ -{{Q ^2} /{2\de ^2}}}
\eqno(2)$$
The uncertainty in the initial position of the pointer produces errors of order
$\de $; when $\de \rightarrow 0$ we recover the
ideal measurement.
Thus,
{\it any} measured value is possible, although large errors are exponentially
suppressed.  There is no mystery in such errors; they are
expected, given the uncertainty associated with the measuring device.
Measurements can even yield
negative values. The
negative values may be unphysical, but they are part of a distribution
representing the measurement of a physical quantity.  They should not be thrown
out, since they give information about the distribution and contribute to the
best estimate of the peak value.  Since
these errors originate in the measuring device, and not in the system under
study, it seems that they cannot depend on any property of the system. However,
closer analysis of these errors
reveals a pattern which clearly reflects
properties of the system under study. The pattern emerges only after selection
of a particular final state of the system.

     Initially, the particle and device are in a product state $\Psi_{in} (x)
\Phi _{in} (Q) $; after the interaction is complete, the state is
$e^{-{i\over \hbar }P~ {{p^{2}} /2m}}
\Psi_{in} (x) \Phi _{in} (Q)$,
in which the particle and the device are correlated. Now we consider kinetic
energy measurements followed by a final measurement of position, with the
particle found far outside the potential well. For the final state we choose a
gaussian wave packet with its center far from the potential well,
$$\Psi _{fin} (x) = (\ep^2 \pi )^{-1/4} e^{-(x-x_{0})^{2}/ 2\ep^2 } ~~~~,
\eqno(3)$$
and we require $\delta > \alpha \hbar^2 /m\de$.
The condition for the particle to be ``far enough" from the potential
well is
$$\alpha {x_0} >>\left( \alpha^2 \hbar^2 /2m \epsilon \right)^2
{}~~~~.\eqno(4)$$
Since $\alpha^2 \hbar^2 /2m = \vert E_0 \vert$, the expression in parentheses
is the ratio of the magnitude of the effect, $\vert E_0 \vert$, to the
precision of the measurement, $\epsilon$.
For more precise measurements of kinetic energy $(\de
\rightarrow 0)$, the final state is selected at increasing distances from the
potential well $(x_0 \rightarrow \infty )$.

     The state of the measuring device after the measurement, and after the
particle is found in the state $\Psi_{fin} (x)$, is obtained by projecting the
correlated state of the particle and measuring device onto the final state of
the particle $\Psi _{fin} (x)$. Apart from normalization, it is
$\Phi _{fin} (Q) = \langle \Psi_{fin} \vert
e^{-{i\over \hbar }P~{{p^{2}} /2m}}
\vert \Psi_{in} \rangle \Phi_{in} (Q)$.
For simplicity, we take $V(x)$
to be a delta-function potential ($a \rightarrow 0$). Then $\Psi_{in} (x)$ is
$\sqrt{\alpha} \exp (-\alpha \vert x\vert)$.  As an integral over $x$,
the final state is
$$
\Phi_{fin} (Q) = \int_{-\infty}^{\infty} dx e^{-(x-x_0 )^2 /2\ep^2 }
e^{-{i\over \hbar} P p^2 /2m}  e^{-\alpha \vert x \vert } \Phi_{in} (Q)
{}~~~~,\eqno(5)$$
up to normalization.  Note that the exponential of $-iPp^2
/2m\hbar$ acts to translate $Q$ in $\Phi_{in} (Q)$.
If we could ignore the part of the integral near $x = 0$,
we could replace $p^2$ with $-\alpha^2$ in Eq. (5),
and the final state of the measuring device would be $\Phi_{fin} (Q) =
\Phi_{in} (Q+\alpha^2 \hbar^2 /2m)$.
We cannot ignore this part of
the integral, but by choosing $x_0$ in $\Psi_{fin} (x)$
to be large, we can suppress it. If we express
$\Psi_{in} (x)$ via its Fourier transform
and replace the operator $p$ with its eigenvalue, we obtain
(up to a normalizing factor)
$$
\Phi_{fin} (Q) = {\pi \over {\hbar \alpha}}
e^{\alpha x_0 - \alpha^2 \delta^2 /2 }
\int dp {{e^{-p^2  \delta^2 /2\hbar^2 - i p x_0/\hbar} }
\over {\alpha^2 \hbar^2 +p^2} }\Phi_{in} (Q - p^2 /2m)
{}~~~~.\eqno(6)$$
This integral has poles at $p= \pm i\alpha \hbar$; we evaluate it by
integration on a contour including a line of $p$ with imaginary part $-i\L$,
for any $\L > \hbar \alpha$.  The integral in \hbox{Eq. (6)}
then reduces to two terms:  a pole term
$$
\Phi_{in} (Q+ {{\alpha^2 \hbar^2}   / {2m}})
{}~~~~, \eqno(7)$$
and a correction term, the integral in Eq. (6)
with $p$ replaced by $p-i\L$.  The pole term represents the measuring device
with its pointer shifted to the negative value $-\alpha^2 \hbar^2 /2m$.
A short computation (see Ref. 4) shows that the correction term can be made
arbitrarily small by taking $x_0$ large, as in Eq. (4).
For $x_0$ large, the final state of the measuring device shows the
``unphysical" result $-\alpha^2 \hbar^2 /2m$ for the kinetic energy, up
to a scatter $\epsilon$ characteristic of the device.

     We thus obtain a correlation between position measurements and prior
kinetic energy measurements:  nearly all particles found far outside the
potential well yielded negative values of kinetic energy.  On the other hand,
we could consider all
particles that produced negative values of kinetic energy, and ask about their
final position. We would find nearly all these particles {\it inside} the well.
The correlation works one way only.  Prior kinetic energy measurements on
particles found far from the well cluster around a negative value, but position
measurements on particles yielding negative values of kinetic energy cluster
around zero.  How do we interpret this one-way correlation?
\goodbreak
{\bf 3. \quad Interpretation.}
\nobreak
Our example suggests that particles in a classically forbidden region have
negative kinetic energy. The conventional interpretation of quantum
mechanics has no place for negative kinetic energy.
However, the conventional interpretation involves an assumption about how
measurements are made.  The conventional interpretation considers measurements
on ensembles of systems prepared in an initial state, without any conditions on
the final state of the systems.  Such an ensemble, defined by initial
conditions only, may be termed a {\it pre-selected} ensemble.  By contrast, we
consider measurements made on {\it pre- and post-selected} ensembles, defined
by both initial and final conditions.  The experiment of the previous section
is an example of a measurement on a pre- and post-selected ensemble.  It is
natural to introduce pre- and post-selected ensembles in quantum theory:  in
the quantum world, unlike the classical world,  complete specification of the
initial state does not determine the final state.

     Also, the measurements we consider are not {\it ideal}.
Real measurements are subject to error. At the same time,
the disturbance they make is bounded.  These two aspects of non-ideal
measurements
go together.  Suppose our measuring device interacts very weakly with the
systems in the ensemble.  We pay a price in precision.  On the other hand, the
measurements hardly disturb the ensemble, and therefore they characterize the
ensemble during the whole intermediate time.  Even non-commuting operators can
be measured at the same time if the measurements are imprecise. When such
measurements are made on pre- and post-selected ensembles, they yield
surprising results.  An operator yields {\it weak} values that need not be
eigenvalues, or even classically allowed.$^{1,6}$  The negative kinetic energy
of the previous section is an example of a weak value.  Another is a
measurable value of 100 for a spin component of a spin-$1/2$ particle.$^1$

     Let us briefly review how weak values arise. The initial wave function of
the measuring device is $\Phi_{in} (Q)$.  After an impulsive measurement of an
operator $C$ on an initial state $\vert a \rangle$
and projection onto a final state $\vert b \rangle$, the final state of the
measuring device is
$\langle b |e^{-iPC/\hbar} |a\rangle \Phi _{in} (Q)
= \sum_i \langle b |c_i \rangle \langle c_i | a \rangle \Phi_{in} (Q-c_i )$.
If $\Phi_{in} (Q)$ is sharply peaked, then the various terms $\Phi_{in} (Q-c_i
)$ will be practically orthogonal.
But suppose $\Phi (Q)$ has a width $\epsilon$. Its Fourier transform has a
width in $P$ of $\hbar /\epsilon$. Small $\vert P \vert$ corresponds to a
measuring device that is coupled weakly to the measured system. If $\epsilon$
is large, then $\vert P \vert$ is small, and we can expand the exponential
$e^{-iPC/\hbar}$ to first order in $P$ to obtain
$\langle b |e^{-iPC/\hbar} |a\rangle \Phi (Q)
\approx  \langle b |1 {-iPC/\hbar} |a\rangle \Phi (Q)
\approx  \langle b |a\rangle e^{-iP C_w /\hbar} \Phi (Q)$.
Here $C_w \equiv {{\langle a |C|b \rangle} / {\langle a|b\rangle}}$
is the {\it weak value} of the operator $C$ for the pre- and post-selected
ensemble defined by $\langle b|$ and $|a\rangle$.

     The definition of a weak value provides us with a new and intuitive
language for describing quantum processes.  In our example, the operators of
total energy $E$, kinetic energy $K$, and potential energy $V$ do not commute.
Therefore, the classical formula $E=K+V$ applies
 only to their expectation values; and the
expectation value of $K$ in any state is positive.  However, the formula
applies to weak values:  $ E _w = K_w +  V_w$,
and the weak value of $K$ is {\it not} necessarily positive.
We know $ E_w =E_0=-\alpha^2 \hbar^2 /2m$, since the
pre-selected state is an energy eigenstate, and $V_w$ vanishes since the
post-selected state is far from the potential well.  Then $K_w = -\alpha^2
\hbar^2 /2m$, the ``unphysical" obtained above in our example!

     In our example, instead of the condition on the initial state of
the measuring device ($\de$ large), we had a condition on the final state of
the particle ($x_0$ large and $\ep > \alpha \hbar^2 /m\de$).
The price is that
we must wait for increasingly rare events.  As measurements of kinetic energy
become more precise ($\epsilon \rightarrow 0$), they disturb the particle more.
To get negative kinetic energies, we must post-select particles further from
the potential well ($x_0 \rightarrow \infty$). As the precision of the
measurement increases, negative kinetic energies become less and less frequent;
in the limit of ideal measurements, the probability vanishes, and so ideal
measurements of kinetic energy never yield negative values.
\goodbreak
{\bf 4. \quad Conclusions.}
\nobreak
{}From the point of view of standard quantum theory, all that we have
produced is a game of errors of measurement. Ideal measurements of kinetic
energy can yield only positive values, since all eigenvalues of the kinetic
energy operator are positive. But in practice, measurements are not exact, and
even if their precision is very good, sometimes -- rarely -- they yield
negative values.  We have seen that if particles are subsequently found far
from the potential well, the measured kinetic energy of these particles comes
out negative.  Consistently, large measurement ``errors" did occur, producing a
distribution peaked at the ``unphysical" negative value $E_0$.

     What special properties of non-ideal measurements led to this result?
First, these measurements involve only bounded disturbances of particle
position. Second, since their precision is limited, they can supply, ``by
error", the necessary negative values. These two properties are intimately
connected: any measurement of kinetic energy causing only bounded changes of
position  must occasionally yield negative values for the kinetic energy.  The
change of $x$ due to the measurement is
$\dot x={i\over {\hbar }}~g(t)~ [ x,~ P~ {p^{2}/ {2m}}]$.
$P$ and $p$ are unchanged during the measurement, so
$x_{fin}-x_{in}=P~ {p/ {m}}$.
{}From here it follows that the change of $x$ is bounded only if the pointer is
in an initial state with $P$ bounded, i.e. if the Fourier transform of
$\Phi_{in} (Q)$ has compact support. But then the support of $\Phi_{in}(Q)$ is
unbounded,$^7$ which immediately implies a nonzero probability for the pointer
to indicate negative values ($Q <0 $). Indeed, the ``game of errors" displays a
remarkable consistency, and this consistency allows negative kinetic energies
to enter physics in a natural way.
The concept of a {\it weak} value of a quantum operator gives precise
meaning to the statement that the kinetic energy of a particle in a classically
forbidden region is negative:  namely, the weak value of the kinetic energy is
negative.
\goodbreak
\bigskip
{\bf \quad Acknowledgement.}
\nobreak
The research was supported by grant 425/91-1
of the the Basic Research Foundation (administered by the Israel Academy of
Sciences and Humanities) and by National Science Foundation grant PHY-8807812.
\goodbreak
\bigskip
\centerline{\bf \quad References}
\medskip
\nobreak
1.  AHARONOV, Y., D. ALBERT \& L. VAIDMAN.  1988. {Phys. Rev. Lett.} {\bf
60}:  1351.

2.  AHARONOV, Y., J. ANANDAN, S. POPESCU \& L. VAIDMAN.  1990.
{Phys. Rev. Lett.} {\bf 64}:  2965.

3.  AHARONOV, Y., S. POPESCU, D. ROHRLICH, \& L. VAIDMAN.  1993. {\it In}
Proc. 4th Int. Symp. Foundations of Quantum Mechanics, Tokyo, 1992.
JJAP Series 9:  251.

4.  AHARONOV, Y., S. POPESCU, D. ROHRLICH \& L. VAIDMAN.  1993.  {Phys.
Rev.} {\bf A48}: 4084.

5.  VON NEUMANN, J. 1983. {Mathematical Foundations of Quantum Theory}
Princeton University Press.  Princeton, New Jersey.

6.  AHARONOV, Y. \& L. VAIDMAN. 1990. {Phys. Rev.} {\bf A41}: 11.
AHARONOV, Y. \& D. ROHRLICH.  1990.  {\it In} {Quantum Coherence} (Proceedings
of the Conference on Fundamental Aspects of Quantum Theory, Columbia,
South Carolina, 1989). J. S. Anandan, Ed. World-Scientific.

7.  If the Fourier transform of $\Phi _{in} (Q)$ has compact
support, then $\Phi_{in} (Q)$ is analytic. The two derivations of
our result, via contour integration and via Taylor expansion
of the exponential in Sec. 3,
both require $\Phi_{in} (Q)$ to be analytic.
\bye